\documentclass[aps,pra,twocolumn, floatfix]{revtex4-1}
\usepackage{graphicx}
\usepackage{verbatim}
\usepackage[fleqn]{amsmath}
\usepackage{bm}
\usepackage{times}
\usepackage[colorlinks,citecolor=blue,linkcolor=red]{hyperref}
\usepackage{color}

\begin{document}
\title{Asymmetric Nonlinear System is Not sufficient for Non-Reciprocal Quantum Wave Diode}
\author{Gaomin Wu}
\affiliation{%
Center for Phononics and Thermal Energy Science, China-EU Joint Center for Nanophononics, Shanghai Key Laboratory of Special Artificial Microstructure Materials and Technology,
School of Physics Sciences and Engineering, Tongji University, Shanghai 200092, China
}%
\author{Yang Long}
\affiliation{%
Center for Phononics and Thermal Energy Science, China-EU Joint Center for Nanophononics, Shanghai Key Laboratory of Special Artificial Microstructure Materials and Technology,
School of Physics Sciences and Engineering, Tongji University, Shanghai 200092, China
}%
\author{Jie Ren}
\email[]{Xonics@tongji.edu.cn}
\affiliation{%
Center for Phononics and Thermal Energy Science, China-EU Joint Center for Nanophononics, Shanghai Key Laboratory of Special Artificial Microstructure Materials and Technology,
School of Physics Sciences and Engineering, Tongji University, Shanghai 200092, China
}%

\date{\today}

\begin{abstract}
We demonstrate symmetric wave propagations in asymmetric nonlinear quantum systems. 
By solving the nonlinear Sch\"ordinger equation, we first analytically prove the existence of symmetric transmission in asymmetric systems with a single nonlinear delta-function interface. We then point out that a finite width of the nonlinear interface region is necessary to produce non-reciprocity in asymmetric systems. 
However, a geometrical resonant condition for breaking non-reciprocal propagation is then identified theoretically and verified numerically. With such a resonant condition,  the nonlinear interface region of finite width behaves like a single nonlinear delta-barrier so that wave propagations in the forward and backward directions are identical under arbitrary incident wave intensity.
As such, reciprocity re-emerges periodically in the asymmetric nonlinear system when changing the  width of interface region.
Finally, similar resonant conditions of discrete nonlinear Sch\"ordinger equation are discussed. 
Therefore, we have identified instances of Reciprocity Theorem that breaking spatial symmetry in nonlinear interface systems is not sufficient to produce non-reciprocal wave propagation.

\end{abstract}


\maketitle

\section{Introduction}

The quest for non-reciprocal wave propagation has spawned vast new designs of rectifiers and diodes in many branches of physics, since it provides the possibility of controlling the energy, information or mass flow. In analogy to electron diodes, there are many theoretical proposals of wave rectifiers to control wave propagation and energy transport. Examples include thermal diodes~\cite{thermal_diodes_exp,thermal_diodes_theory_1,thermal_diodes_theory_2,thermal_diodes_theory_3,thermal_diodes_theory_4,phononic_diodes_expandth, negdiff2013} that are capable of controlling thermal heat transfer in nonreciprocal phononic systems; 
spin Seebeck diodes~\cite{predictedrec2013, theoryasy2013, nanospin2013} that can rectify pure spin current by temperature bias;
acoustics diodes~\cite{nonlinear1,liang2009,popa2014}  with potential applications in manipulating vibrational energy for control of destruction and uni-directional sonic barrier for energy harvesting;
and optical diodes or isolators~\cite{optical_diodes_exp,optical_diodes_th_1,optical_diodes_th_nonlinear_ag} to suppress undesired light interference in laser and high-density integrated optical circuits. Some of them have been verified experimentally~\cite{experiment_acoustics_diodes,optical_diodes_exp,phononic_diodes_expandth,thermal_diodes_exp}.

The definition of non-reciprocal wave propagation is that: the transmitted power at the same incident amplitude and frequency is sensibly different in two opposite propagation directions~\cite{sufficient1}. 
To obtain the non-reciprocity~\cite{Rayleigh} in linear systems, the time-reversal symmetry should be broken. For instance, Faraday effect is applied in optical isolators to break time-reversal symmetry with the application of magneto-acoustic materials~\cite{mag_acoustic}.
The other way to achieve wave non-reciprocity without breaking time-reversal symmetry is to consider nonlinearity, such as non-reciprocal acoustic devices using nonlinear medium~\cite{nonlinear1,liang2009}, nonlinear electronic circuit~\cite{popa2014} for frequency conversion, 
nonlinear optical photonic crystals~\cite{nonlinear_photonic}, and 
thermal rectifiers using nonlinear lattices~\cite{thermal_diodes_theory_1,thermal_diodes_theory_2}.

It has been widely and well accepted that although nonlinearity or spatial asymmetry alone can not guarantee the non-reciprocity, both of them together is sufficient to provide nonreciprocal wave propagation~\cite{sufficient1,sufficient2,sufficient3}. However, we should reminder that this interpretation has never been proved strictly, and can be regarded as a hypothesis.
And in this paper, we will demonstrate that this interpretation is flawed and invalid, {\it i.e.}, 
asymmetric nonlinear system is not sufficient for non-reciprocal quantum wave diode!

In this work, we tackle the issue with one-dimensional nonlinear quantum structure described by nonlinear Schr\"odinger equation (NLSE) with spatially varying coefficients, which is often used to describe nonlinear models especially for propagation of solitons~\cite{NLSE}.
In Section~\ref{sectionI}, we examine the problem of asymmetric wave propagation with plane waves passing across nonlinear $\delta$-function potential. We will demonstrate that, when the nonlinearity appears only at a single spot, the wave propagation forward will always be identical to propagation backward. 
In Section~\ref{sectionII}, we build a model where we place nonlinearity at two spots to form a finite width interface (scattering) region and find that non-reciprocity exists in this case. But, by changing the width of the interface (scattering) region bounded by two nonlinear potential spots, or equivalently say, by changing the distance of two nonlinear potential spots, reciprocal wave propagation would appear again in the asymmetric nonlinear system, when a resonant condition is satisfied. At this condition, two nonlinear spot potentials are effectively equivalent to a single nonlinear spot potential. 
We also show that similar effects of this finite width nonlinear interface region can also be offered by a finite width interface region bounded by a single-point nonlinearity and a linear interface.
In Section~\ref{sectionIII}, we build a discrete layer model to analyze the transport in discrete systems, by using the discrete nonlinear Schr\"odinger equation. We will show the existence of similar resonant conditions, in different discrete language, but revealing the same physical mechanism.

We prove a general theorem, as a consequence, that the observation of non-reciprocity must always imply not only in both nonlinearity and spatial asymmetry but also with taking the geometrical properties into consideration: here, the width of nonlinear interface region. 
Given the geometry-tuned resonant conditions satisfied, reciprocity re-emerges in a structure with even both spatial asymmetry and nonlinearity.

\section{Results and Discussions}
\subsection{Single nonlinear $\delta$-function potential}\label{sectionI}
\begin{figure}
\begin{center}
\includegraphics[width=0.33\textwidth,clip]{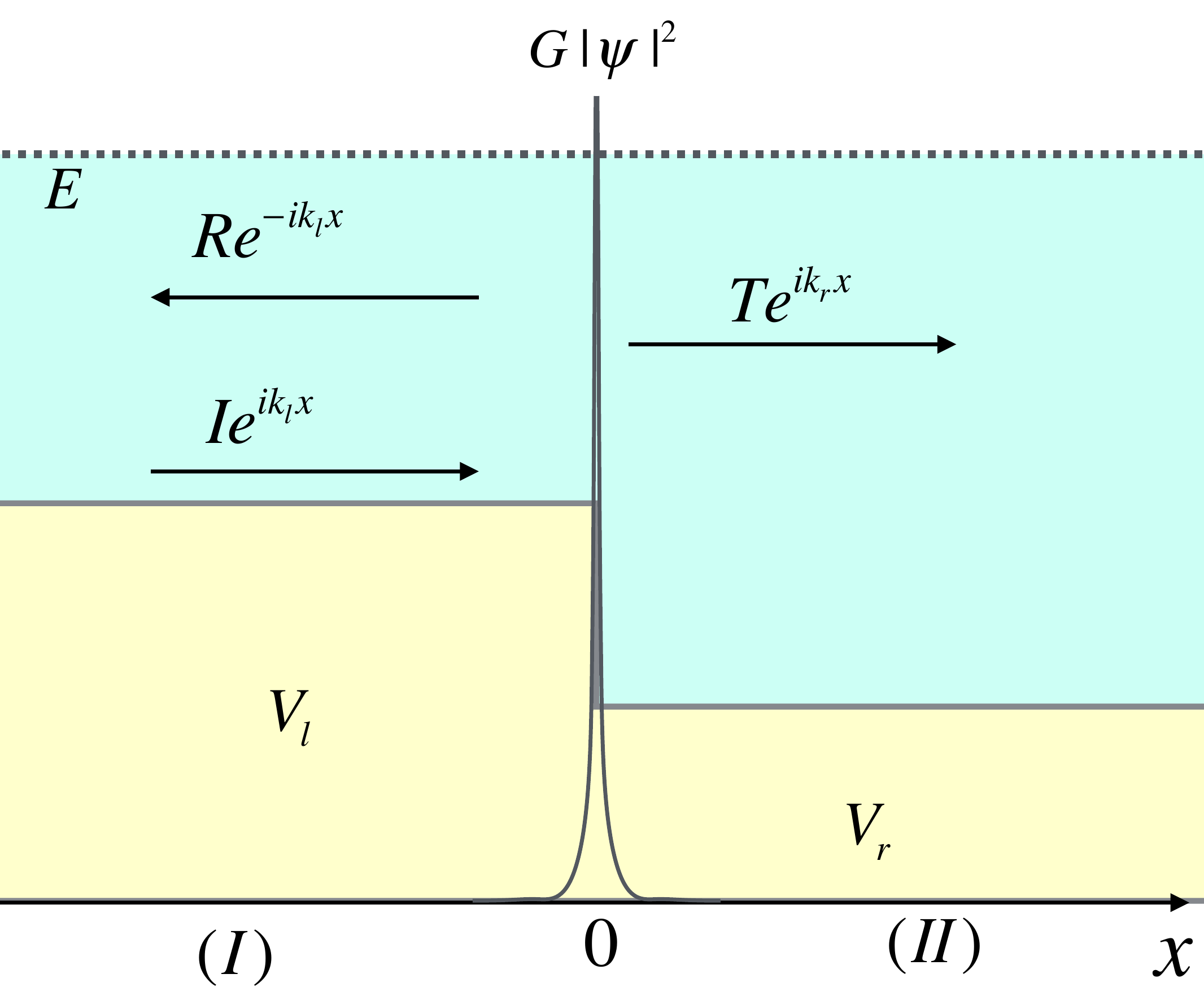}
\vspace{-5mm}
\caption{Geometry of a single layer nonlinear interface. A plane wave of amplitude $I$
strikes a nonlinear  $\delta$-function potential,
giving rise to a reflected wave of amplitude $R$ and transmitted wave of amplitude $T$,
where $k_l$ and $k_r$ are determined by $V_{l(r)}$ in region ($I$) and ($II$), respectively, described by equation 
$k_{l(r)}=\sqrt{{2m(E-V_{l(r)})}}/{\hbar}$. We consider scattering states with $E>V$. The backward wave propagation can be similarly described.}
\label{fig1}
\end{center}
\end{figure}

We first consider the problem of plane quantum wave propagating through a very thin nonlinear interface 
at the origin.
To construct this very thin interface layer, a single nonlinear Dirac $\delta$-function potential is located
at $x=0$ as described by:
\begin{equation}
-\frac{\hbar^2}{2m}\left[\frac{d^2}{dx^2}- G\delta(x)\left|\Psi(x)\right|^2\right] \Psi(x) = \left(E-V(x)\right) \Psi(x), 
\label{NLS}
\end{equation}
where $G$ denote the nonlinear strength of the $\delta$-function potential, and $V(x)$ is the linear potential.
Here, we consider plane wave that enable us to analytically derive the transmitted coefficient 
and rectifying factor for the forward and backward wave propagations.
Potential of this problem is depicted in Fig.~\ref{fig1}, with asymmetric potential $V(x)=V_l$ when $x<0$, and $V(x)=V_r$ when $x>0$.

Solving Eq.(\ref{NLS}) will yield the forward transmission coefficient $t_f$ 
(where subscript f denotes forward) from left to right as the function of transmitted 
wave amplitude$|T|^2$:
\begin{equation}\label{transco}
t_f\equiv\frac{|T|^2k_r}{|I|^2k_l}=\frac{4k_rk_l}{(k_l+k_r)^2+G^2|T|^4}
\end{equation}
The ratio $k_r/k_l$ in the definition of transmission coefficient is to normalize it since $V_l\not=V_r$. Without this normalization, forward transmission coefficient might exceed unity if $V_l>V_r$, because of the existence of gain in the system. The conservation of probability current in this case can be verified as 
$k_r|T|^2 = k_l|I|^2-k_l|R|^2$.
Note that we can get the backward transmission coefficient $t_b$ in this problem by exchanging 
$k_l$ and $k_r $, same as to reverse the model. Obviously, backward transmission coefficient is identical to the forward coefficient 
in this problem.
Therefore, spatial asymmetry is not sufficient to give rise to non-reciprocity in nonlinear systems, if the nonlinear system is constructed by a single nonlinear $\delta$-function potential.

\subsection{Two nonlinear $\delta$-function potential}\label{sectionII}
\begin{figure}
\begin{center}
\includegraphics[width=0.44\textwidth,clip]{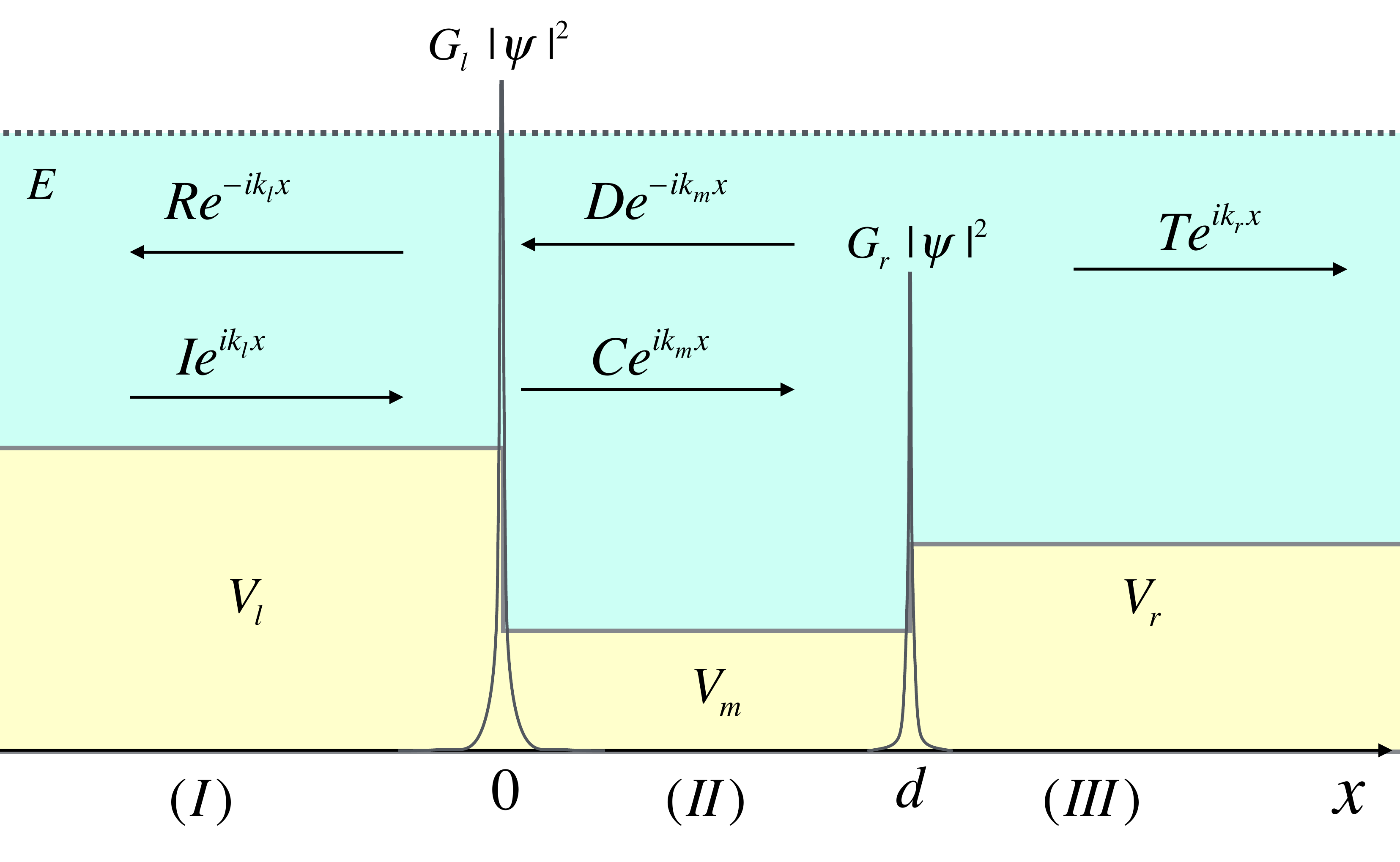}
\vspace{-7mm}
\caption{Geometry of a finite width nonlinear interface region, bounded by two nonlinear $\delta$-function potentials. In the middle interface region
(II), there are a reflected wave of amplitude $D$ and a transmitted wave of amplitude $C$, 
where $k_m$ is determined by $V_m$ (the linear potential in region (II)). 
$k_{l,m,r}=\sqrt{{2m(E-V_{l,m,r})}}/{\hbar}$. The backward wave propagation can be similarly described.
}
\label{fig2}
\end{center}
\end{figure}

Let us now consider to add one more thin nonlinear layer 
in the distance $d$ to the origin. 
The nonlinear Schr\"odinger equation for this problem is 
\begin{eqnarray}
-\frac{\hbar^2}{2m}&&\left[\frac{d^2}{dx^2}-G_l\delta(x)|\Psi(x)|^2-G_r\delta(x-d)|\Psi(x)|^2\right] \Psi(x)
 \nonumber \\
 && = (E-V(x))\Psi(x),
\label{2NLS}
\end{eqnarray}
where $G_r$ and $G_l$ denote the nonlinear strength of right and left delta-function potentials 
($G_r, G_l>0$ for the delta barriers), respectively.
And $V_{l,m,r}$ are the linear potentials in three different regions divided by the two barriers, respectively. Geometry of this setup 
is depicted in Fig.~\ref{fig2}, with potential $V(x)=V_l$ at $x<0$, $V(x)=V_m$ at $0<x<d$, and $V(x)=V_r$ at $x>d$.

The transmission coefficient can be derived analytically as following, 
considering $E>V$.
The general solution in the region $(I)$ at $x<0$ is
\begin{eqnarray}
\Psi(x)=Ie^{ik_lx}+Re^{-ik_lx}.
\label{fL}
\end{eqnarray}
Similarly, in the region$(II)$ at $0<x<d$, 
\begin{eqnarray}
\Psi(x)=Ce^{ik_mx}+De^{-ik_mx},
\label{fM}
\end{eqnarray}
and in the region$(III)$ at $x>d$, 
\begin{eqnarray}
\Psi(x)=Te^{ik_rx},
\label{fR}
\end{eqnarray}
as we consider scattering from the left.

The continuity of $\Psi(x)$ at $x = 0$ and $x = d$ requires that
\begin{eqnarray}\label{1boundary}
I+R &=& C+D, \\
Te^{ik_rd} &=& Ce^{ik_md}+De^{-ik_md}.
\end{eqnarray}
To figure out the relationship of derivatives at the interface, let us integrate Eq.(\ref{2NLS})
from $-\epsilon$ to $+\epsilon$, and take the limit $\epsilon \rightarrow 0$: 
\begin{eqnarray}
\int_{-\epsilon}^{+\epsilon}\frac{d^2}{dx^2}\Psi(x)dx-\int_{-\epsilon}^{+\epsilon}G_l\delta(x)|\Psi(x)|^2\Psi(x)\big]dx\nonumber\\
 = \int_{-\epsilon}^{+\epsilon}(E-V(x))\Psi(x)dx.
\end{eqnarray}
Since the last integral vanishes in the limit $\epsilon \rightarrow 0$, so the second boundary condition yields:
\begin{eqnarray}\label{2boundary1}
\frac{d\Psi}{dx}\bigg|_{+\epsilon}-\frac{d\Psi}{dx}\bigg|_{-\epsilon}
=G_l|\Psi(0)|^2\Psi(0).
\end{eqnarray}
Similarly, at $x = d$
\begin{eqnarray}\label{2boundary2}
\frac{d\Psi}{dx}\bigg|_{d+\epsilon}-\frac{d\Psi}{dx}\bigg|_{d-\epsilon}
=G_r|\Psi(d)|^2\Psi(d).
\end{eqnarray}
Taking derivatives of $\Psi(x)$ at three regions [see Eqs.~(\ref{fL}, \ref{fM}, \ref{fR})],  
and substituting them into Eqs.~(\ref{2boundary1}) and (\ref{2boundary2}),  thus the second boundary conditions read:
\begin{eqnarray}\label{bound1}
ik_l(I-R)
-G_l|C+D|^2(C+D)=ik_m(C-D).
\end{eqnarray}
\begin{eqnarray}\label{bound2}
ik_m(Ce^{ik_md}-De^{-ik_md})=(ik_r+G_r|T|^2)Te^{ik_rd}.
\end{eqnarray}
Note that $\Psi(x=0)=C+D$ and $\Psi(x=d)=Te^{ik_rd}$.

By combining Eq.~(\ref{1boundary}) and Eq.~(\ref{bound2}), we can describe $C$ and $D$ as a function of $T$:
\begin{eqnarray}\label{CD}
C &= \frac{1}{2}(1+N_1-iN_2)Te^{i(k_r-k_m)d}.\nonumber\\
D &=  \frac{1}{2}(1-N_1+iN_2)Te^{i(k_r+k_m)d}.
\end{eqnarray}
with $N_1 = k_r/k_m, N_2 = {G_r|T|^2}/{k_m}$. 
Then we can derive from Eq.(\ref{CD}) that  
\begin{eqnarray}
|C+D|^2 =\frac{N_1^2 + N_2^2+ 1}{2}|T|^2,
\end{eqnarray}
and note that the width of interface region $d$ in Eq.(\ref{CD}) does not exist in $|C+D|^2$.
Combining Eqs.~(\ref{1boundary}) and (\ref{bound1}), we can describe $I$ as a function of $C$ and $D$:
\begin{eqnarray}
I = \frac{1}{2}C(1+N_3-iN_4)+\frac{1}{2}D(1-N_3-iN_4).
\end{eqnarray}
with $N_3 = k_m/k_l$, $N_4 = G_l|C+D|^2/k_l=G_l(N_1^2 + N_2^2+ 1)|T|^2/(2k_l)$. 
After regrouping and substitution we can obtain:
\begin{eqnarray}\label{result}
8\frac{|I|^2}{|T|^2} &=& (M_1^2+M_2^2+M_3^2+M_4^2)\nonumber\\
&+&(M_1^2-M_2^2-M_3^2+M_4^2)\cos(2k_md)\nonumber\\
&-&2(M_1M_3-M_2M_4)\sin(2k_md),
\end{eqnarray}
with
\begin{eqnarray}
M_1 &=& 1+N_1N_3, \nonumber\\
M_2 &=& N_1+N_3-N_2N_4, \nonumber\\
M_3 &=& N_1N_4+N_2, \nonumber\\
M_4 &=& N_4+N_2N_3.
\end{eqnarray}
Eq.~(\ref{result}) contains eight parameters $-$ 
the amplitude of incident wave $I$, 
 the amplitude of transmitted wave $T$, width of the nonlinear interface region $d$, nonlinear strength of two delta-function potential $G_l$ and $G_r$, and wave number in three regions $k_l$, $k_m$, and $k_r$.
The forward transmission coefficient $t_f$ can thus be derived analytically by the definition $t_f\equiv\frac{|T|^2k_r}{|I|^2k_l}$. For the reversed direction, the backward transmission coefficient can be obtained by exchanging $G_r\leftrightarrow G_l$ and $k_r\leftrightarrow k_l$, which is just to reverse the model.

\begin{figure}
\begin{center}
\includegraphics[width=0.5\textwidth]{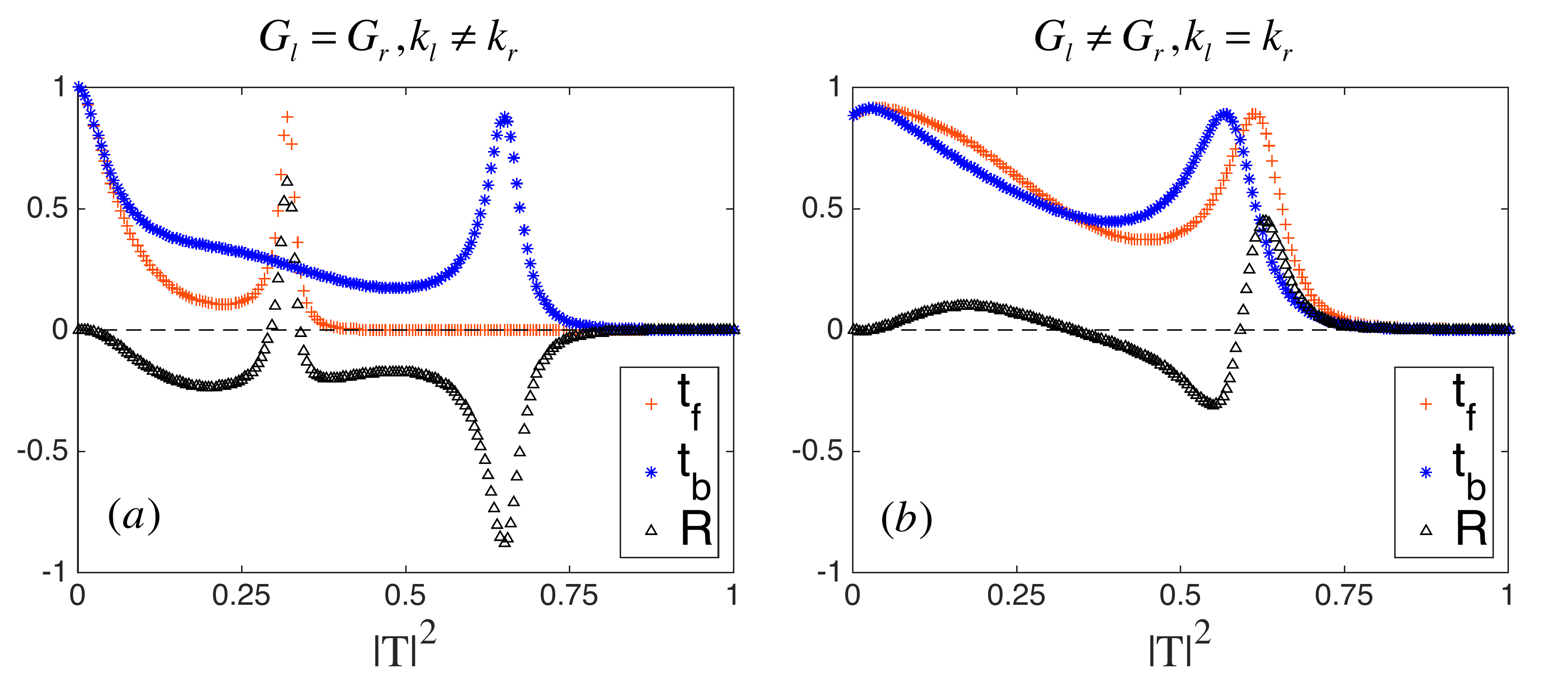}
\vspace{-9mm}
\caption{(a) Asymmetric nonlinear strength where $G_l=1$ and $G_r=3$
 and identical linear potential in three region where
 $k_l=k_m=k_r=0.1$, $d=0.08\pi/k_m$.
 (b) Symmetric nonlinear strength where $G_l=G_r=1$ 
 and asymmetry from different linear potential in three region where
 $k_l=0.2$ and $k_r=0.1$  and $k_m=0.09$, $d=0.09\pi/k_m$.}
\label{fig3}
\end{center}
\end{figure}

Obviously,  Eq.~(\ref{result})  will change after left-right exchange with subscript $l\leftrightarrow r$, which implies that transmission coefficient in this model is direction-dependent so that non-reciprocal. Numerical results can be calculated to verify the non-reciprocity in a clearer way.
To quantify the efficiency of the non-reciprocity, we define the rectifying factor
\begin{equation}
R = t_f - t_b,
\end{equation}
which shows non-reciprocity with non-zero value and with value of $\pm 1$ 
when approaches maximal non-reciprocity.
Fig.~\ref{fig3} illustrates numerical results 
with different parameters. $t_f$, $t_b$ and $R$ are plotted as function of transmitted wave intensities $|T|^2$. As a result, we can see significant non-reciprocal wave propagation in Fig.~\ref{fig3}.

By comparing the reciprocal model in Section~\ref{sectionI} and the non-reciprocal model in Section~\ref{sectionII}, we would like to sum up that non-reciprocal wave propagation is obtained as the results of three factors:\\
(1) Nonlinearity, in this case provided by nonlinear $\delta$-function potential; \\
(2) Spatial asymmetry, provided by either difference between $G_l$ and $G_r$ or $k_l$ and $k_r$;\\
(3) A width of the interface layer, indicated by $d$.

According to the analytical results of transmission coefficient Eq.~(\ref{result}), 
the length of interface region $d$ appears only in the term $\cos(2k_md)$ and $\sin(2k_md)$, since terms$N_1,N_2,N_3$ and $N_4$ contain no $d$ . 
Obviously, changing the width of interface region yields a periodical change
of transmission coefficient. 
When $2k_md=2n\pi, n=0,1,2...$, the situation would be identical to that of $d=0$.
Substituting $\cos(2k_md)=1$ and $\sin(2k_md)=0$ into Eq.~(\ref{result}), the forward transmission coefficient 
can be analytically derived as:
\begin{equation}
t_f=\frac{4k_rk_l}{(k_l+k_r)^2+(G_l+G_r)^2|T|^4}.
\end{equation}
When $d=0$, two barriers 
have no distance and combine as a single nonlinear $\delta$-function potential with overlapped nonlinear coefficient
$G=G_l+G_r$, where non-reciprocity does not show up under any intensity of transmission waves. 
It means that, even though non-reciprocity is obtained by satisfying the three factors 
mentioned above, when you move the barrier at $x=d$, non-reciprocity vanishes periodically when 
$d = n\pi/k_m$. Width of the interface region should avoid the these points to produce non-reciprocity.
Tuning $k_m$ will give the similar effect, but here we focus on tuning the interface region width $d$.

\begin{figure}
\begin{center}
\includegraphics[width=0.4\textwidth]{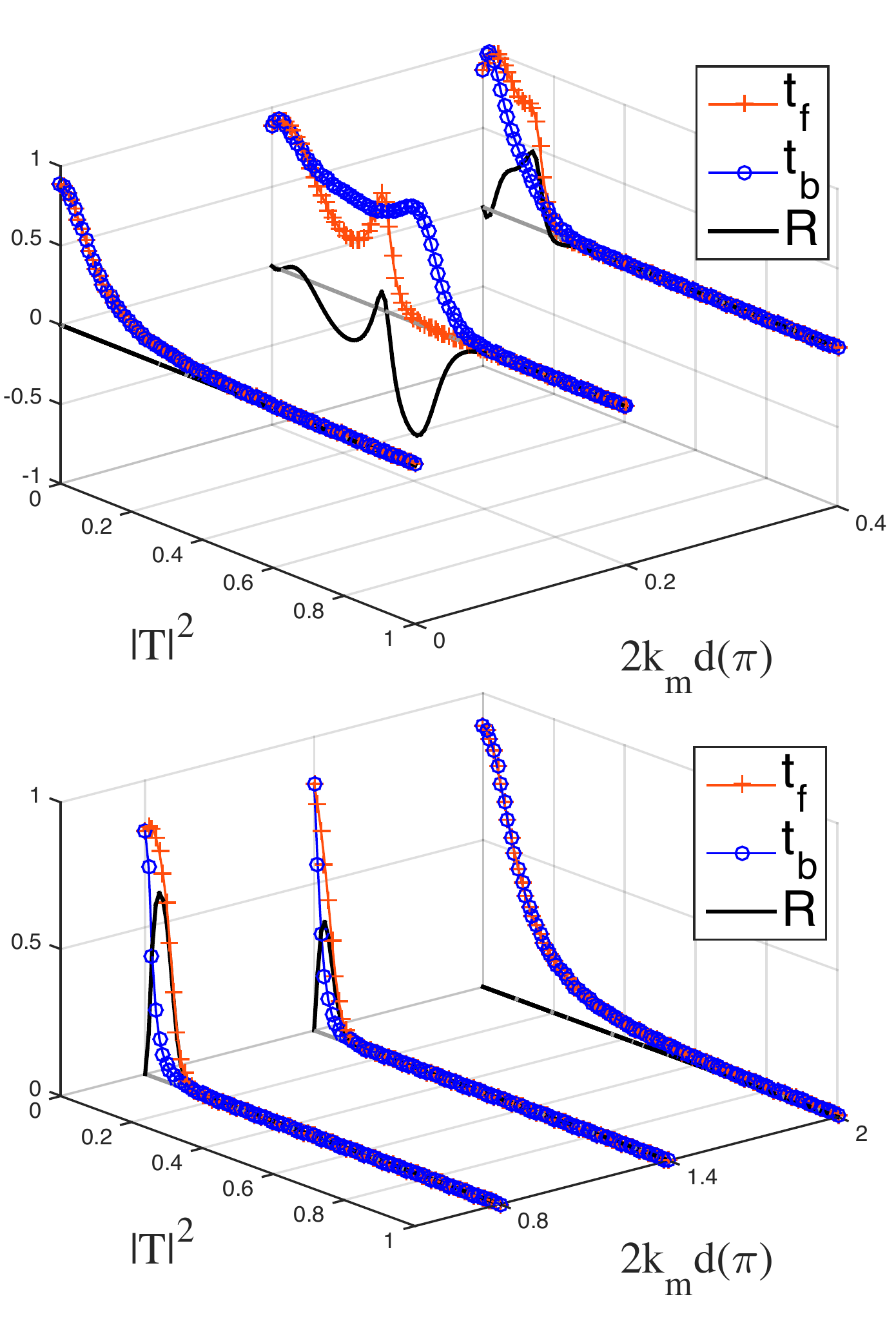}
\vspace{-5mm}
\caption{The forward and backward transmission coefficient and rectifying factor, denoted by $t_f, t_b, R$ respectively,  as functions of the intensity of transmitted wave $|T|^2$ and $k_md$ with unit of $\pi$.}
\label{fig4}
\end{center}
\end{figure}

We analyze a case where $G_l=2,G_r=1$, $k_l=0.2$ and $k_r=0.1$  and $k_m=0.09$ and set $2k_md\in[0, 2\pi]$, a period of the transformation.
As shown in Fig.~\ref{fig4},  when $k_md=0$, the transmission is symmetry under direction reversal indicting the reciprocal transport; When $k_md$ is small, the forward transmission coefficient $t_f$ and backward $t_b$ reach the peak at different intensity of transmitted 
 wave $|T|^2$. As $k_md$ is tuned to be greater in one period, the peaks of two transmission coefficients 
 gradually overlap near the zero transmitted wave intensity. Then, at the resonant condition where $2k_md$ reaches $2\pi$, two functions coincide and the rectifying factor $R$ is exactly zero which means the situations of wave propagating forward and wave propagation backward are identical under any intensity of incident wave. Thus, a symmetric wave transportation emerges even with nonlinearity and spatial symmetry breaking.

Moreover, we also find similar phenomena in an interface with nonlinear potential 
only at one spot followed by a linear constant potential $V_m$ layer. The width of the interface is $d$
and still it is geometrically asymmetry.
This model can be easily built by setting $G_r=0$. It can be easily proved by taking $G_r=0$ into Eq.~(\ref{result}) that as long as the potential of the middle part $V_m$ is different from the adjoining potential $V_r$ (The case of $V_m=V_r$ is identical to the case of single nonlinear $\delta$-function layer), the wave propagation is still nonreciprocal. The asymmetric wave propagation in this problem is reasonable, since the three factors that lead to non-reciprocity are satisfied. Asymmetric design of the interface satisfies the need of asymmetry. 
The transmission coefficient still changes periodically with width of the linear interface $d$ and 
non-reciprocity vanishes periodically by changing the term $2k_md$. 

\subsection{Two nonlinear layer described by DNLS}
\label{sectionIII}

Dirac $\delta$ function is often used as an approximation of a thin layer. In Section~\ref{sectionI} and Section~\ref{sectionII}, the nonlinear layer is represented by the product of the probability of the particle $|\Psi|^2$ and $\delta$ function. More recent works of approximating nonlinearity in thin layers have been demonstrated with discrete nonlinear Schr\"odinger (DNLS) equation~\cite{DNLS}, which is considered to be a reasonable approximation of layered phononic and photonic crystals~\cite{DNLS_USE}. Coefficients in the DNLS equation can represent different strength of nonlinearity of each layers. 
Therefore, in this section, we will use DNLS to re-do the job in Section~\ref{sectionI} and Section~\ref{sectionII}, yet in discrete model. 
We will prove a similar resonant condition analogy to that of the former continuous system.

\begin{figure}[ht]
\begin{center}
\includegraphics[width=0.4\textwidth]{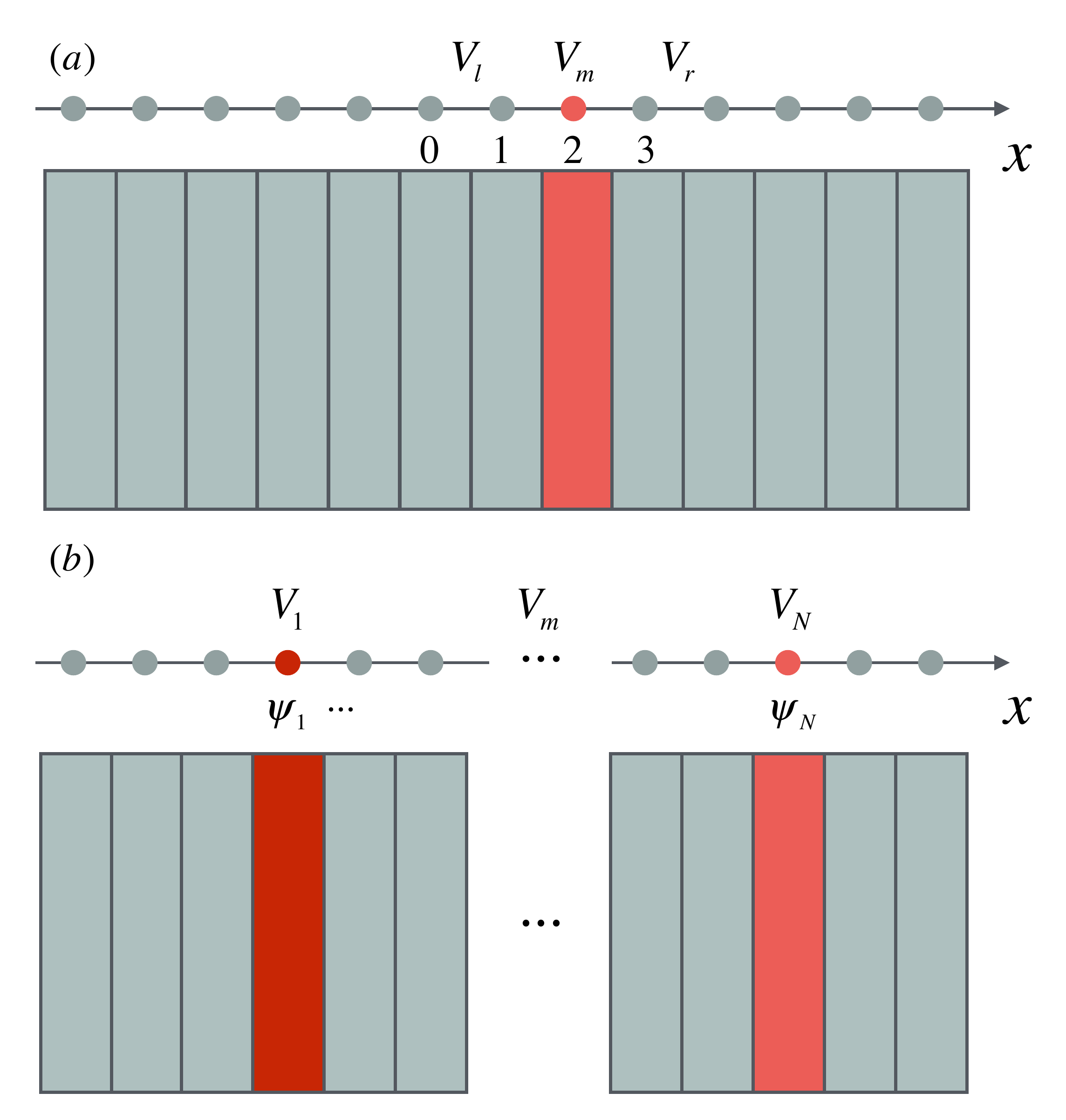}
\vspace{-5mm}
\caption{Geometry of discrete nonlinear systems. (a) The dark (red) site represents the single nonlinear thin layer sandwiched by linear sites described by gray sites. (b) The two dark (red) sites represent two nonlinear layers. Two nonlinear layers have different nonlinear strength to satisfy asymmetric geometry. Each site has corresponding on-site energy described by $V_n$.}
\label{fig5}
\end{center}
\end{figure}

Figure~\ref{fig5} shows sketches of the discrete nonlinear model that can be described by the stationary DNLS equation in one dimension:
\begin{equation}
(E - V_n)\psi_n = 2\psi_n 
-\psi_{n+1} - \psi_{n-1} + G_n |\psi_n|^2 \psi_n,
\label{dnls}
\end{equation}
where $V_n$ denote the on-site energy of site $n$.
Fig.~\ref{fig5}(a) is the discrete form of model in Section~\ref{sectionI}, and similarly Fig.~\ref{fig5}(b) is the equivalent to the continuous model in Section~\ref{sectionII}.

In the single nonlinearity model, we will look for the analytical solution to the  transmission problem by similar scattering approach presented by Refs.~\cite{sufficient1, sufficient2}.
\begin{eqnarray}\label{algebra1}
&\psi_0 = I+R,\quad\psi_1&= Ie^{ik_l} + Re^{-ik_l},   \nonumber\\
&\psi_2 = T,\quad\quad\psi_{3}&= Te^{ik_r}.
\end{eqnarray}
By solving the DNLS equation set of this model
\begin{eqnarray}\label{DNLS1}
&\psi_{1} &= (2-E+V_m+G|\psi_2|^2)\psi_2-\psi_{3}, \nonumber\\
&\psi_0 &= (2-E-V_l)\psi_{1}-\psi_{2},
\end{eqnarray}
we can obtain the analytical result of  transmission coefficient.
The amplitude of incident wave $I$ can be derived from Eq.~(\ref{algebra1}):
\begin{equation}\label{incident}
I = \frac{\psi_0e^{-ik_l}-\psi_1}{e^{-ik_l}-e^{ik_l}}.
\end{equation}

Now let us calculate the square of modulus of the numerator in Eq.~(\ref{incident}) and remember we are going to represent $I$ as a function of $T$.
Derived from DNLS Eq.~(\ref{DNLS1}), $\psi_1$ and $\psi_0$ can be represented by amplitude of transmission wave $T$ as 
\begin{eqnarray}
&\psi_{1} &= (2-E+V_m+G|T|^2-e^{ik_r})T,   \nonumber\\
&\psi_0 &= ((2-E-V_l)(2-E+V_m+G|T|^2-e^{ik_r})-1)T.   \nonumber
\end{eqnarray}
Thus the square of modulus of the numerator in Eq.~(\ref{incident}) is
\begin{equation}
\label{fe}
|\psi_0e^{-ik_l}-\psi_1|^2= |2-E+V_m+G|T|^2-e^{ik_r}-e^{ik_l}|^2|T|^2,
\end{equation} 
where we use the dispersion relation $2-E+V_l = 2\cos(k_l)$.
By Combining Eq.~(\ref{fe}) and Eq.~(\ref{incident}), transmission coefficient can be obtained as 
\begin{eqnarray}\label{transco_dis}
t_f &=& \frac{\sin{k_r}|T|^2}{\sin{k_l}|I|^2} \nonumber\\ 
&=& \frac{4\sin{k_l}\sin{k_r}}{|2-E+V_m+G|T|^2-e^{ik_r}-e^{ik_l}|^2}.   
\end{eqnarray}
Similar to the ratio $k_r/k_l$ in Section~\ref{sectionII}, the ratio $\sin{k_r}/\sin{k_l}$ appears to properly define the transmission $t_f$. This analytical result of transmission coefficient shows reasonable reciprocity that $t_b=t_f$ under exchanging $l\leftrightarrow r$, which verifies the statement we mentioned earlier that a single nonlinear thin layer with spatial asymmetry cannot guarantee non-reciprocal wave transportation.
This result leads us to consider resonant condition that can merge two nonlinear layers into a single one, which will yield reciprocity as one single nonlinearity does.

In the second discrete model shown in Fig.~\ref{fig5}(b), we assume that the linear sites sandwiched by two nonlinear layers have the same on-site energy $V_m$, where $m$ denote middle, {\it i.e.}, $V_n = V_m (1<n<N)$. The other two parts of the model are linear structures with $V_n=V_1 (n\leq1)$ and $V_n=V_N (n\geq N)$.
Similarly,
\begin{eqnarray}\label{algebra2}
\psi_0 = I+R,\quad\psi_1 &=& Ie^{ik_l} + Re^{-ik_l},  \nonumber\\
\psi_N = T,\quad\psi_{N+1} &=& Te^{ik_r}.
\end{eqnarray}
By taking Eq.~(\ref{algebra2}) into equation set of DNLS that describe the model:
\begin{eqnarray}
&\psi_{N-1} &= (2-E+V_N+G_r|\psi_N|^2)\psi_N-\psi_{N+1}\nonumber\\
&\psi_{N-2} &= (2-E+V_m)\psi_{N-1}-\psi_{N}\nonumber\\
&&\quad\quad\quad\quad...\nonumber\\
&\psi_1 &= (2-E+V_m)\psi_{2}-\psi_{3}\nonumber\\
&\psi_{0} &= (2-E+V_1+G_l|\psi_1|^2)\psi_1-\psi_{2}
\end{eqnarray}
The array of DNLS have $N$ equations, and the $N-2$ equations that having exactly the same recursive pattern describe only the middle linear part. Thus, we can find recursive relationship to finally represent $\psi_1$ as a function of $\psi_N$ and $\psi_{N-1}$. 
Consider  
\begin{eqnarray}
\psi_1 = \mu_n\psi_{n-1} + \upsilon_n\psi_n, 
\end{eqnarray}
with $n=3, 4, ..., N-1, N$, where sequence $\{\mu_n\}$ and $\{\upsilon_n\}$ start at $n = 3$ with $\mu_3 = (2 -E+V_m)$, $\upsilon_3 = -1$.
With simple recursion steps, recursive formula of $\{\mu_n\}$ and $\{\upsilon_n\}$ can be obtained as
\begin{eqnarray}
\bigg(\begin{matrix}\mu_{n+1} \\\upsilon_{n+1} \end{matrix} \bigg)
= \bigg(\begin{matrix} {2-E+V_m}&1\\{-1}&0\end{matrix} \bigg)
\bigg(\begin{matrix}\mu_{n}\\\upsilon_{n} \end{matrix} \bigg).
\end{eqnarray}

In the section above, in order to achieve reciprocity in system with geometric asymmetry and nonlinearity, we need to reach the resonant condition where two nonlinear layers are merged into one.
In this case, it means $\psi_1 = \pm\psi_N$, the minus sign occurs because the definition of transmission coefficient contains only $|\psi|^2$. 
Thus, reciprocity re-emerges when $\mu_N=0$ and $\upsilon_N=\pm1$ are satisfied in $\psi_1 = \mu_N\psi_{N-1} + \upsilon_N\psi_N$. Therefore, we identify the resonant condition in this discrete nonlinear model  as
\begin{equation} \label{Dresonant}
\bigg(\begin{matrix}\mu_N \\\upsilon_N \end{matrix} \bigg)=\bigg(\begin{matrix} {2-E+V_m}&1\\{-1}&0\end{matrix} \bigg)^{N-3}
\bigg(\begin{matrix}{2-E+V_m}\\{-1} \end{matrix} \bigg)=\bigg( \begin{matrix}0\\\pm1\end{matrix}\bigg).
\end{equation}
At this resonant condition, we obtain the reciprocal transmission coefficient  as 
\begin{equation}\label{transco_dis2}
t_f = \frac{4\sin{k_l}\sin{k_r}}{|2-E+\alpha+(G_r+G_l)|T|^2-e^{ik_r}-e^{ik_l}|^2},
\end{equation}
where $\alpha =V_1+V_N-V_m$. 
The transmission coefficient under resonant condition is actually equivalent to the one of single nonlinearity model. If we replace the sum of the relative on-site energy on nonlinear sites $(V_1+V_N-V_m)$ in Eq.~(\ref{transco_dis2}) with $V_m$ and replace the sum of nonlinear strength at two nonlinear sites $G_r+G_l$ with $G$, Eq.~(\ref{transco_dis2}) can be converted into Eq.~(\ref{transco_dis}). Clearly, the transmission is symmetric under left-right exchange, when satisfying the resonant condition Eq.~(\ref{Dresonant}), otherwise non-reciprocal. 

\section{Conclusions}
In summary, we have rechecked the conventional interpretation that spatial asymmetry is sufficient to produce non-reciprocal wave propagation in nonlinear quantum systems, and have shown that this state is incorrect. We have shown the necessity to take geometrical properties - the width of nonlinear interface into concern and have discussed resonant conditions where reciprocity re-emerges in an otherwise non-reciprocal system.
 
Considering the continuous model, we have found three sufficient factors to give rise to non-reciprocity in nonlinear systems: 1) nonlinearity; 2) spatial asymmetry; 3) finite width of the interface scattering region. We have touched upon the specific role of the width of the interface scattering region in producing non-reciprocity: when the width $d$ meets the resonant condition $2k_md=2n\pi, n=0, 1, 2,...$, two nonlinearities are added and behave as a single nonlinear spot so that non-reciprocity vanishes. Hence, the width of the interface scattering region is an essential factor to the phenomenology reported herein. Similarly, we have identified similar resonant conditions in the discrete model described by DNLS.
 
Factors that affect the resonant condition are little bit different in two forms of models. In continuous model, the width of the layer denoted by $d$ affects the resonant condition, while in discrete model it is the number of sites between two nonlinear sites. However, it is important to point out that the resonant condition also depends on the energy of the incident wave $E$ and the potentials, in both continuous and discrete models. It means that if the nonlinear structure is fixed, only particular frequencies (energies) of wave will be able to transport reciprocally. Emergence of symmetric quantum transport in asymmetric nonlinear quantum systems is not found here for the first time. Previous studies in the heat diode and spin Seebeck diode have shown that even in the strong asymmetric interface, i.e., Fermi-Boson coupling system, the nonlinear quantum transport can be symmetric under particular conditions~\cite{negdiff2013, predictedrec2013}.

Our present results suggest some interesting potential exploration for future works. For example, to investigate the existence of resonant condition in the (un)periodic model of linear structures sandwiched by multiple nonlinear layers that effectively merges all the nonlinear layers into one. This renders us the possibility to construct a special layered nonlinear phononic or photonic crystal periodically inserted with linear crystals that can give rise to reciprocity for selected frequencies of waves, and otherwise non-reciprocal.


\acknowledgements
This work is supported by the NSFC with grant No. 11775159, the National Youth 1000 Talents Program in China, and the startup Grant at Tongji University.


\begin{thebibliography}{300}
\expandafter\ifx\csname natexlab\endcsname\relax\def\natexlab#1{#1}\fi
\expandafter\ifx\csname bibnamefont\endcsname\relax
  \def\bibnamefont#1{#1}\fi
\expandafter\ifx\csname bibfnamefont\endcsname\relax
  \def\bibfnamefont#1{#1}\fi
\expandafter\ifx\csname citenamefont\endcsname\relax
  \def\citenamefont#1{#1}\fi
\expandafter\ifx\csname url\endcsname\relax
  \def\url#1{\texttt{#1}}\fi
\expandafter\ifx\csname urlprefix\endcsname\relax\def\urlprefix{URL }\fi
\providecommand{\bibinfo}[2]{#2}
\providecommand{\eprint}[2][]{\url{#2}}

\bibitem[{\citenamefont{Chang et~al.}(2006)\citenamefont{Chang, Okawa,
  Majumdar, and Zettl}}]{thermal_diodes_exp}
\bibinfo{author}{\bibfnamefont{C.~W.} \bibnamefont{Chang}},
  \bibinfo{author}{\bibfnamefont{D.}~\bibnamefont{Okawa}},
  \bibinfo{author}{\bibfnamefont{A.}~\bibnamefont{Majumdar}}, \bibnamefont{and}
  \bibinfo{author}{\bibfnamefont{A.}~\bibnamefont{Zettl}},
  \bibinfo{journal}{Science} \textbf{\bibinfo{volume}{314}},
  \bibinfo{pages}{1121} (\bibinfo{year}{2006}).

\bibitem[{\citenamefont{Terraneo et~al.}(2002)\citenamefont{Terraneo, Peyrard,
  and Casati}}]{thermal_diodes_theory_1}
\bibinfo{author}{\bibfnamefont{M.}~\bibnamefont{Terraneo}},
  \bibinfo{author}{\bibfnamefont{M.}~\bibnamefont{Peyrard}}, \bibnamefont{and}
  \bibinfo{author}{\bibfnamefont{G.}~\bibnamefont{Casati}},
  \bibinfo{journal}{Physical Review Letters} \textbf{\bibinfo{volume}{88}},
  \bibinfo{pages}{094302} (\bibinfo{year}{2002}).

\bibitem[{\citenamefont{Li et~al.}(2004)\citenamefont{Li, Wang, and
  Casati}}]{thermal_diodes_theory_2}
\bibinfo{author}{\bibfnamefont{B.}~\bibnamefont{Li}},
  \bibinfo{author}{\bibfnamefont{L.}~\bibnamefont{Wang}}, \bibnamefont{and}
  \bibinfo{author}{\bibfnamefont{G.}~\bibnamefont{Casati}},
  \bibinfo{journal}{Physical Review Letters} \textbf{\bibinfo{volume}{93}},
  \bibinfo{pages}{184301} (\bibinfo{year}{2004}).

\bibitem[{\citenamefont{Casati}(2005)}]{thermal_diodes_theory_3}
\bibinfo{author}{\bibfnamefont{G.}~\bibnamefont{Casati}},
  \bibinfo{journal}{Chaos: An Interdisciplinary Journal of Nonlinear Science}
  \textbf{\bibinfo{volume}{15}}, \bibinfo{pages}{015120}
  (\bibinfo{year}{2005}).

\bibitem[{\citenamefont{Segal and Nitzan}(2005)}]{thermal_diodes_theory_4}
\bibinfo{author}{\bibfnamefont{D.}~\bibnamefont{Segal}} \bibnamefont{and}
  \bibinfo{author}{\bibfnamefont{A.}~\bibnamefont{Nitzan}},
  \bibinfo{journal}{Physical Review Letters} \textbf{\bibinfo{volume}{94}},
  \bibinfo{pages}{034301} (\bibinfo{year}{2005}).

\bibitem[{\citenamefont{Li et~al.}(2012)\citenamefont{Li, Ren, Wang, Zhang,
  H{\"a}nggi, and Li}}]{phononic_diodes_expandth}
\bibinfo{author}{\bibfnamefont{N.}~\bibnamefont{Li}},
  \bibinfo{author}{\bibfnamefont{J.}~\bibnamefont{Ren}},
  \bibinfo{author}{\bibfnamefont{L.}~\bibnamefont{Wang}},
  \bibinfo{author}{\bibfnamefont{G.}~\bibnamefont{Zhang}},
  \bibinfo{author}{\bibfnamefont{P.}~\bibnamefont{H{\"a}nggi}},
  \bibnamefont{and} \bibinfo{author}{\bibfnamefont{B.}~\bibnamefont{Li}},
  \bibinfo{journal}{Reviews of Modern Physics} \textbf{\bibinfo{volume}{84}},
  \bibinfo{pages}{1045} (\bibinfo{year}{2012}).

\bibitem[{\citenamefont{Ren and Zhu}(2013{\natexlab{a}})}]{negdiff2013}
\bibinfo{author}{\bibfnamefont{J.}~\bibnamefont{Ren}} \bibnamefont{and}
  \bibinfo{author}{\bibfnamefont{J.~X.} \bibnamefont{Zhu}},
  \bibinfo{journal}{Physical Review B} \textbf{\bibinfo{volume}{87}},
  \bibinfo{pages}{241412} (\bibinfo{year}{2013}{\natexlab{a}}).

\bibitem[{\citenamefont{Ren}(2013)}]{predictedrec2013}
\bibinfo{author}{\bibfnamefont{J.}~\bibnamefont{Ren}},
  \bibinfo{journal}{Physical Review B} \textbf{\bibinfo{volume}{88}},
  \bibinfo{pages}{220406} (\bibinfo{year}{2013}).

\bibitem[{\citenamefont{Ren and Zhu}(2013{\natexlab{b}})}]{theoryasy2013}
\bibinfo{author}{\bibfnamefont{J.}~\bibnamefont{Ren}} \bibnamefont{and}
  \bibinfo{author}{\bibfnamefont{J.~X.} \bibnamefont{Zhu}},
  \bibinfo{journal}{Physical Review B} \textbf{\bibinfo{volume}{88}},
  \bibinfo{pages}{094427} (\bibinfo{year}{2013}{\natexlab{b}}).

\bibitem[{\citenamefont{Ren et~al.}(2014)\citenamefont{Ren, Fransson, and
  Zhu}}]{nanospin2013}
\bibinfo{author}{\bibfnamefont{J.}~\bibnamefont{Ren}},
  \bibinfo{author}{\bibfnamefont{J.}~\bibnamefont{Fransson}}, \bibnamefont{and}
  \bibinfo{author}{\bibfnamefont{J.~X.} \bibnamefont{Zhu}},
  \bibinfo{journal}{Physical Review B} \textbf{\bibinfo{volume}{89}},
  \bibinfo{pages}{214407} (\bibinfo{year}{2014}).

\bibitem[{\citenamefont{Liu et~al.}(2015)\citenamefont{Liu, Du, Sun, Gao, and
  Guo}}]{nonlinear1}
\bibinfo{author}{\bibfnamefont{C.}~\bibnamefont{Liu}},
  \bibinfo{author}{\bibfnamefont{Z.}~\bibnamefont{Du}},
  \bibinfo{author}{\bibfnamefont{Z.}~\bibnamefont{Sun}},
  \bibinfo{author}{\bibfnamefont{H.}~\bibnamefont{Gao}}, \bibnamefont{and}
  \bibinfo{author}{\bibfnamefont{X.}~\bibnamefont{Guo}},
  \bibinfo{journal}{Physical Review Applied} \textbf{\bibinfo{volume}{3}},
  \bibinfo{pages}{064014} (\bibinfo{year}{2015}).

\bibitem[{\citenamefont{Liang et~al.}(2009)\citenamefont{Liang, Yuan, and
  Cheng}}]{liang2009}
\bibinfo{author}{\bibfnamefont{B.}~\bibnamefont{Liang}},
  \bibinfo{author}{\bibfnamefont{B.}~\bibnamefont{Yuan}}, \bibnamefont{and}
  \bibinfo{author}{\bibfnamefont{J.~C.} \bibnamefont{Cheng}},
  \bibinfo{journal}{Physical Review Letters} \textbf{\bibinfo{volume}{103}},
  \bibinfo{pages}{104301} (\bibinfo{year}{2009}).

\bibitem[{\citenamefont{Popa and Cummer}(2014)}]{popa2014}
\bibinfo{author}{\bibfnamefont{B.~I.} \bibnamefont{Popa}} \bibnamefont{and}
  \bibinfo{author}{\bibfnamefont{S.~A.} \bibnamefont{Cummer}},
  \bibinfo{journal}{Nature Communications} \textbf{\bibinfo{volume}{5}},
  \bibinfo{pages}{3398} (\bibinfo{year}{2014}).

\bibitem[{\citenamefont{Hu et~al.}(2011)\citenamefont{Hu, Li, Zhang, Yang,
  Gong, and Zhang}}]{optical_diodes_exp}
\bibinfo{author}{\bibfnamefont{X.}~\bibnamefont{Hu}},
  \bibinfo{author}{\bibfnamefont{Z.}~\bibnamefont{Li}},
  \bibinfo{author}{\bibfnamefont{J.}~\bibnamefont{Zhang}},
  \bibinfo{author}{\bibfnamefont{H.}~\bibnamefont{Yang}},
  \bibinfo{author}{\bibfnamefont{Q.}~\bibnamefont{Gong}}, \bibnamefont{and}
  \bibinfo{author}{\bibfnamefont{X.}~\bibnamefont{Zhang}},
  \bibinfo{journal}{Advanced Functional Materials}
  \textbf{\bibinfo{volume}{21}}, \bibinfo{pages}{1803} (\bibinfo{year}{2011}).

\bibitem[{\citenamefont{Tocci et~al.}(1995)\citenamefont{Tocci, Bloemer,
  Scalora, Dowling, and Bowden}}]{optical_diodes_th_1}
\bibinfo{author}{\bibfnamefont{M.~D.} \bibnamefont{Tocci}},
  \bibinfo{author}{\bibfnamefont{M.~J.} \bibnamefont{Bloemer}},
  \bibinfo{author}{\bibfnamefont{M.}~\bibnamefont{Scalora}},
  \bibinfo{author}{\bibfnamefont{J.~P.} \bibnamefont{Dowling}},
  \bibnamefont{and} \bibinfo{author}{\bibfnamefont{C.~M.}
  \bibnamefont{Bowden}}, \bibinfo{journal}{Applied Physics Letters}
  \textbf{\bibinfo{volume}{66}}, \bibinfo{pages}{2324} (\bibinfo{year}{1995}).

\bibitem[{\citenamefont{Peng et~al.}(2014)\citenamefont{Peng, Li, and
  She}}]{optical_diodes_th_nonlinear_ag}
\bibinfo{author}{\bibfnamefont{N.}~\bibnamefont{Peng}},
  \bibinfo{author}{\bibfnamefont{X.}~\bibnamefont{Li}}, \bibnamefont{and}
  \bibinfo{author}{\bibfnamefont{W.}~\bibnamefont{She}},
  \bibinfo{journal}{Optics Express} \textbf{\bibinfo{volume}{22}},
  \bibinfo{pages}{17546} (\bibinfo{year}{2014}).

\bibitem[{\citenamefont{Liang et~al.}(2010)\citenamefont{Liang, Guo, Tu, Zhang,
  and Cheng}}]{experiment_acoustics_diodes}
\bibinfo{author}{\bibfnamefont{B.}~\bibnamefont{Liang}},
  \bibinfo{author}{\bibfnamefont{X.~S.} \bibnamefont{Guo}},
  \bibinfo{author}{\bibfnamefont{J.}~\bibnamefont{Tu}},
  \bibinfo{author}{\bibfnamefont{D.}~\bibnamefont{Zhang}}, \bibnamefont{and}
  \bibinfo{author}{\bibfnamefont{J.~C.} \bibnamefont{Cheng}},
  \bibinfo{journal}{Nat Mater} \textbf{\bibinfo{volume}{9}},
  \bibinfo{pages}{989} (\bibinfo{year}{2010}).

\bibitem[{\citenamefont{Lepri and Casati}(2011)}]{sufficient1}
\bibinfo{author}{\bibfnamefont{S.}~\bibnamefont{Lepri}} \bibnamefont{and}
  \bibinfo{author}{\bibfnamefont{G.}~\bibnamefont{Casati}},
  \bibinfo{journal}{Physical Review Letters} \textbf{\bibinfo{volume}{106}},
  \bibinfo{pages}{164101} (\bibinfo{year}{2011}).

\bibitem[{\citenamefont{Rayleigh and Lindsay}(1945)}]{Rayleigh}
\bibinfo{author}{\bibfnamefont{B.}~\bibnamefont{Rayleigh}, \bibfnamefont{John
  William~Strutt}} \bibnamefont{and} \bibinfo{author}{\bibfnamefont{R.~B.}
  \bibnamefont{Lindsay}}, \emph{\bibinfo{title}{The theory of sound}}
  (\bibinfo{publisher}{Dover Publications}, \bibinfo{year}{1945}).

\bibitem[{\citenamefont{L{\"u}thi}(2007)}]{mag_acoustic}
\bibinfo{author}{\bibfnamefont{B.}~\bibnamefont{L{\"u}thi}},
  \bibinfo{journal}{Solid State Sciences} \textbf{\bibinfo{volume}{65}},
  \bibinfo{pages}{13} (\bibinfo{year}{2007}).

\bibitem[{\citenamefont{Gallo et~al.}(2001)\citenamefont{Gallo, Assanto,
  Parameswaran, and Fejer}}]{nonlinear_photonic}
\bibinfo{author}{\bibfnamefont{K.}~\bibnamefont{Gallo}},
  \bibinfo{author}{\bibfnamefont{G.}~\bibnamefont{Assanto}},
  \bibinfo{author}{\bibfnamefont{K.~R.} \bibnamefont{Parameswaran}},
  \bibnamefont{and} \bibinfo{author}{\bibfnamefont{M.~M.} \bibnamefont{Fejer}},
  \bibinfo{journal}{Applied Physics Letters} \textbf{\bibinfo{volume}{79}},
  \bibinfo{pages}{314} (\bibinfo{year}{2001}).

\bibitem[{\citenamefont{Li and Ren}(2014)}]{sufficient2}
\bibinfo{author}{\bibfnamefont{N.}~\bibnamefont{Li}} \bibnamefont{and}
  \bibinfo{author}{\bibfnamefont{J.}~\bibnamefont{Ren}},
  \bibinfo{journal}{Scientific Reports} \textbf{\bibinfo{volume}{4}},
  \bibinfo{pages}{6228} (\bibinfo{year}{2014}).

\bibitem[{\citenamefont{D'Ambroise et~al.}(2012)\citenamefont{D'Ambroise,
  Kevrekidis, and Lepri}}]{sufficient3}
\bibinfo{author}{\bibfnamefont{J.}~\bibnamefont{D'Ambroise}},
  \bibinfo{author}{\bibfnamefont{P.~G.} \bibnamefont{Kevrekidis}},
  \bibnamefont{and} \bibinfo{author}{\bibfnamefont{S.}~\bibnamefont{Lepri}},
  \bibinfo{journal}{Journal of Physics A Mathematical Theoretical}
  \textbf{\bibinfo{volume}{45}}, \bibinfo{pages}{2077} (\bibinfo{year}{2012}).

\bibitem[{\citenamefont{Mihalache et~al.}(1993)\citenamefont{Mihalache, Torner,
  Moldoveanu, Panoiu, and Truta}}]{NLSE}
\bibinfo{author}{\bibfnamefont{D.}~\bibnamefont{Mihalache}},
  \bibinfo{author}{\bibfnamefont{L.}~\bibnamefont{Torner}},
  \bibinfo{author}{\bibfnamefont{F.}~\bibnamefont{Moldoveanu}},
  \bibinfo{author}{\bibfnamefont{N.~C.} \bibnamefont{Panoiu}},
  \bibnamefont{and} \bibinfo{author}{\bibfnamefont{N.}~\bibnamefont{Truta}},
  \bibinfo{journal}{Journal of Physics A: Mathematical and General}
  \textbf{\bibinfo{volume}{26}}, \bibinfo{pages}{L757} (\bibinfo{year}{1993}).

\bibitem[{\citenamefont{Eilbeck et~al.}(1985)\citenamefont{Eilbeck, Lomdahl,
  and Scott}}]{DNLS}
\bibinfo{author}{\bibfnamefont{J.~C.} \bibnamefont{Eilbeck}},
  \bibinfo{author}{\bibfnamefont{P.~S.} \bibnamefont{Lomdahl}},
  \bibnamefont{and} \bibinfo{author}{\bibfnamefont{A.~C.} \bibnamefont{Scott}},
  \bibinfo{journal}{Physica D: Nonlinear Phenomena}
  \textbf{\bibinfo{volume}{16}}, \bibinfo{pages}{318} (\bibinfo{year}{1985}).

\bibitem[{\citenamefont{Kosevich and Mamalui}(2002)}]{DNLS_USE}
\bibinfo{author}{\bibfnamefont{A.~M.} \bibnamefont{Kosevich}} \bibnamefont{and}
  \bibinfo{author}{\bibfnamefont{M.~A.} \bibnamefont{Mamalui}},
  \bibinfo{journal}{Journal of Experimental and Theoretical Physics}
  \textbf{\bibinfo{volume}{95}}, \bibinfo{pages}{777} (\bibinfo{year}{2002}).

\end{thebibliography}


\end{document}